\documentclass[aps,prb,twocolumn,showpacs,superscriptaddress,floatfix,citeautoscript]{revtex4}
\usepackage{graphicx}
\usepackage{times}

\begin{document}

\title{
Interaction Effects in the Mesoscopic Regime: \\
A Quantum Monte Carlo Study of Irregular Quantum Dots
}

\author{Amit Ghosal}
\affiliation{Department of Physics, Duke University,
Durham, North Carolina 27708-0305}

\author{C. J. Umrigar}
\affiliation{Theory Center and Laboratory of Atomic and Solid State Physics, Cornell University,
Ithaca, New York 14853}

\author{Hong Jiang}
\thanks{Current address: Institut f\"ur Theoretische Physik, J.W.Goethe- Universit\"at, Frankfurt am Main, Germany.}
\affiliation{Department of Physics, Duke University,
Durham, North Carolina 27708-0305}
\affiliation{Department of Chemistry, Duke University,
Durham, North Carolina 27708-0354}

\author{Denis Ullmo}
\thanks{Permanent address: Laboratoire de Physique Th\'eorique et
Mod\`eles Statistiques (LPTMS), 91405 Orsay Cedex, France}
\affiliation{Department of Physics, Duke University,
Durham, North Carolina 27708-0305}

\author{Harold U.~Baranger}
\affiliation{Department of Physics, Duke University,
Durham, North Carolina 27708-0305}

\date{\today}

\begin{abstract}
We address the issue of accurately treating interaction effects in the
mesoscopic regime by investigating the ground state properties of isolated
irregular quantum dots. Quantum Monte Carlo techniques are used to calculate
the distributions of ground state spin and addition energy. We find a
reduced probability of high spin and a somewhat larger even/odd alternation
in the addition energy from quantum Monte Carlo than in local spin density
functional theory. In both approaches, the even/odd effect gets smaller with
increasing number of electrons, contrary to the theoretical understanding of
large dots. We argue that the local spin density approximation over predicts
the effects of interactions in quantum dots.

\end{abstract}


                                                                                
\maketitle


The interplay between correlations and quantum mechanical interference of
electron states, long at center stage in condensed matter physics, has been
traditionally investigated in disordered systems but can also be probed in
confined systems, such as quantum dots \cite{Kouwenhoven97,ABG}.
In these latter, the confinement leads to mesoscopic fluctuations
\cite{Kouwenhoven97,Folk01} which in turn modify the role of Coulomb
repulsion between electrons within the dots. Quantum dots offer the great
practical advantage of experimental tunability in the study of this
interplay \cite{Kouwenhoven97,Folk01}.

Quantum dots of different size give rise naturally to different descriptions.
For \textit{small} dots, including both vertical \cite{vertdotexpt} and few
electron lateral \cite{latdotexpt} dots, circular symmetry is preserved and
plays a critical role. In this limit quantum Monte Carlo (QMC) calculations
have been performed \cite{Bolton96,Shumway00,Pederiva00,Nieminen02,Guo03}
as well as simpler density functional theory (DFT) simulations
\cite{wingreen,Reimann02RMP}. A comparative study for the weak interaction
regime \cite{Shumway00,Pederiva00} confirmed the validity of the DFT method
in this small dot limit.

For \textit{large irregular} dots \cite{Kouwenhoven97,Folk01}, on the other
hand, all spatial symmetries are broken.  For a sufficiently irregular shape,
the motion of electrons within the dot is chaotic, which then
justifies modeling the single-particle energies and wave functions by random
matrix theory and random plane waves, respectively \cite{Bohigas91,Berry77}.
Furthermore, interaction effects in these larger dots are often treated within
the random phase approximation (RPA)
for gas parameter $r_s$ (the ratio between the interaction energy and kinetic
energy, formally defined as $r_s=1/\sqrt{\pi n} a_0$ for 2D bulk systems,
which thus identifies the strength of the interaction) of order 1 or
smaller \cite{Blanter,ABG}.  The ``universal Hamiltonian" picture
\cite{KurlandPRB00,ABG} that emerges leads to statistical predictions for
various quantities, such as the ground state spin or addition energy
\cite{denis,gonzalo}. Once temperature is taken into account,
\cite{gonzalo} these are in good agreement with experimental data \cite{Sivan96,MarcusPatel98,EnsslinLuscher01,EnsslinLindemann02}. One notable
feature is a substantial difference at zero temperature between dots containing
an odd number of electrons, $N$, and those in which this number is even.
Experiments have not to date performed at a sufficiently low temperature to
probe this feature.
This even/odd effect persists for large $N$
with an essentially unchanged magnitude provided that $r_s$ remains constant.

In order to go beyond statistical predictions and address features of
individual irregular dots, an approach which accurately treats the combination
of mesoscopic fluctuations and interaction effects is needed.  DFT appeared
a natural choice for such studies, and, indeed, microscopic calculations of
ground state energies for large irregular dots ($N \!\sim\! 200$) were carried
out within the framework of the local spin density approximation (LSDA)
\cite{Jiang03,Jiang04,denis3}.
The statistics of the LSDA results turned out, however, to be in qualitative
disagreement with the earlier predictions, even for the modest interaction
strengths ($r_s \!\sim\! 1.5$) that are experimentally relevant; for
instance, in the LSDA results at zero temperature, the even/odd effect is
nearly absent. In fact, there were several indications of stronger
interactions in LSDA than those obtained from RPA. The striking discrepancy
between the two approaches -- both of which are believed to be valid in the
range of $r_s$ considered -- combined with the absence of experimental
statistics for low temperature keeps this problem open.

Here we take up the issue of accurately treating interaction effects in the
mesoscopic regime. We consider irregular quantum dots with up to 30 electrons. 
The lack of symmetry-induced shell structure makes irregular dots
qualitatively different from circular dots; in particular, the mesoscopic
interference effects are both more subtle and more generic. In this regime,
where the universal Hamiltonian picture is not expected to hold because of
the modest size, we use QMC to treat the interactions much more carefully
than in LSDA. To this end, we present QMC calculations of the addition energy
and ground state spin for such dots, and compare to corresponding LSDA results.


We consider a model quantum dot consisting of electrons moving in a two
dimensional plane, with kinetic energy $(-\frac{1}{2}\sum_i \nabla_i^2)$,
and interacting with each other by long-range Coulomb repulsion
($\sum_{i<j}|{\bf r}_i - {\bf r}_j|^{-1}$). All energies are expressed in
atomic units, defined by $\hbar \!=\! e^2/\epsilon \!=\! m^* \!=\! 1$,
with electronic
charge $e$, effective mass $m^*$, and dielectric constant of the medium
$\epsilon$.  The electrons are confined by an external (quartic) potential
   \begin{equation}
V_{\rm ext}(x,y) = a\left[\frac{x^4}{b}+b y^4 -2\lambda x^2 y^2 +
                           \gamma (x-y) x y r\right]
\label {eq:iqo}
   \end{equation}
where $r \!=\! \sqrt{x^2+y^2}$.
This simple form of $V_{\rm ext}$ breaks all symmetries except time reversal
invariance.
It leads to chaotic motion of the electrons inside the dot \cite{denis_PR},
which is the experimental situation for large dots;
\cite{Kouwenhoven97,Sivan96,MarcusPatel98,EnsslinLuscher01,EnsslinLindemann02}
in fact, not only
the bare $V_{\rm ext}$ but also the self-consistent potential leads to chaotic
dynamics \cite{Jiang03,Jiang04}.
We have studied potential (\ref{eq:iqo}) for a
range of parameters and report here results for $a=0.002$ (which controls
$r_s$), $b=\pi/4$, $\gamma$ between $0.1$ and $0.2$ (break spatial symmetries),
and $\lambda$ between $0.53$ and $0.67$.  For these parameters,
the dynamics in the bare potential is chaotic. We accumulate statistics for
six dots formed by different sets of parameters ($\lambda$ and $\gamma$) from
the above range. We study dots with $N \!=\! 10$ to $30$ electrons which
yields a range $r_s \!=\! 1.8$ to $1.3$\cite{foot2}.

Variational (VMC) and diffusion (DMC) Monte Carlo techniques\cite{Foulkes01RMP}
were used to calculate the energies $E(N,S)$ of our model quantum dots for each
$N$ and spin $S$. We investigated $S=0$, $1$, and $2$ for even $N$, and
$S=1/2$, $3/2$, and $5/2$ for odd $N$. For a given $V_{\rm ext}$, the ground
state energy $E_{\rm GS}$ and the ground state spin $S_{\rm GS}$ were
determined for each $N$.

The trial wave function used in QMC, $\Psi_T$, is written as a linear
combination of products of up- and down-spin Slater determinants multiplied
by a Jastrow factor.
Each Slater determinant is constructed from single-particle Kohn-Sham (KS)
orbitals obtained using the LSDA functional.
The Jastrow factor effectively describes the dynamic correlation
between the electrons coming from their mutual repulsion,
whereas the near-degeneracy or static correlation is taken into account by
having more than one determinant.
We optimize the Jastrow parameters and determinant coefficients by minimizing
the variance of the local energy \cite{Umrigar88PRL}.

In a second stage, we use fixed-node DMC \cite{Umrigar93JCP,Foulkes01RMP}
to project the optimized many-body wavefunction onto a better approximation of
the true ground state.  The fixed-node DMC energy is an upper bound to the
true energy and depends only on the nodes of the trial wave function, i.e.,
only on the linear combination of determinants. (The Jastrow factor affects
the statistical error of the energy but not its expectation value.)
The statistical error in the energy $E_{\rm DMC}(N,S)$ obtained in this way is
smaller than the single-particle mean level spacing $\Delta$ by about two
orders of magnitude and hence is insignificant.
The systematic error from the fixed-node approximation of the many-body
wave function is, however, difficult to estimate though experience
suggests it is often small.
We have included Slater determinants for which the sum of the KS
single-particle energies are up to $\Delta$ greater than the sum of the
KS single-particle energies for the ground state KS determinant.
This amounts to taking mostly one and sometimes two or three Slater
determinants in the $\Psi_T$ expansion. Increasing the energy window
from $\Delta$ to $3\Delta$ failed to reduce $E_{\rm DMC}$ although
it sometimes reduced $E_{\rm VMC}$.

\begin{figure}
\includegraphics[width=3in,clip]{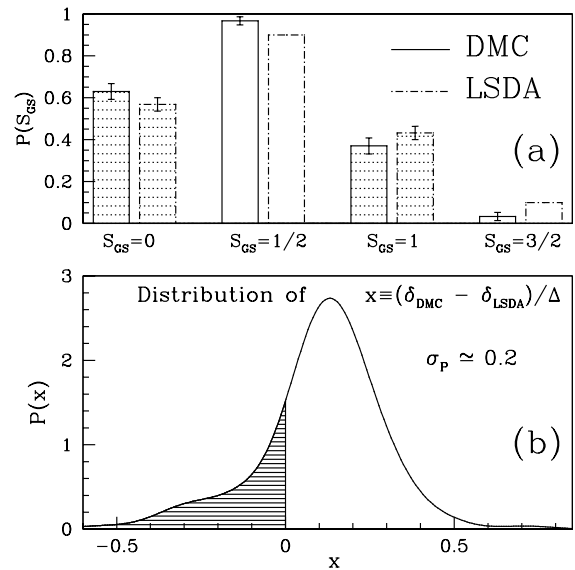}
\caption{(a) Distribution of $S_{\rm GS}$ from DMC and LSDA calculations.
Shaded histograms are for dots with even $N$ while unfilled bars are for
odd $N$; data is collected for $N \!=\! 10$-$30$ and six realizations of
$V_{\rm ext}$, which are also used to estimate the statistical error.
Though the differences are small, LSDA predicts a larger probability of
non-trivial $S_{\rm GS}$.
(b) Distribution of the difference in ``spin gap'' obtained using DMC and
LSDA, normalized by the mean level spacing $\Delta$. The large width of
the distribution ($\sigma_P$) indicates a significant difference between
the two techniques. Note that $x$ (defined in the figure) is primarily
positive; negative values occur predominantly when the ground state has
non-trivial spin.
LSDA is, therefore, making non-trivial spin states more probable by lowering
their energies compared to DMC results. [A sliding Gaussian window of width
0.08 is used to give a smooth estimate for $P(x)$.]
}
\label{fig:Fig1}
\end{figure}

We present the distribution of $S_{\rm GS}$ obtained from both DMC and LSDA
\cite{foot3}
in Fig.~1(a). Within a model of effectively non-interacting electrons,
$S_{\rm GS}$ is $0$ or $1/2$ for even $N$ and odd $N$, respectively, due to
standard up/down filling of the orbitals.  We see that the probability of
finding a non-trivial $S_{\rm GS}$, i.e.\ not zero or half, is substantial.
Interestingly, this probability is reduced in DMC
calculations compared to LSDA.  Note that the differences between the two
distributions, although clearly visible, are not large and are therefore not
much bigger than the statistical error given the relatively small data set
($21 \times 6 = 126$ cases total).
There are, however, significant correlations between the LSDA and DMC results.
In fact,
$S_{\rm GS}$ from DMC is, up to one exception, {\em always the same as or lower
than} that from LSDA.  As a consequence the statistical error on
$\langle S_{\rm GS}^{\rm LSDA}-S_{\rm GS}^{\rm DMC} \rangle$ is only 30\% of
its value. Thus, there is a clear difference between LSDA and DMC in the
predicted ground state spin.

\begin{figure}
\includegraphics[width=3in,clip]{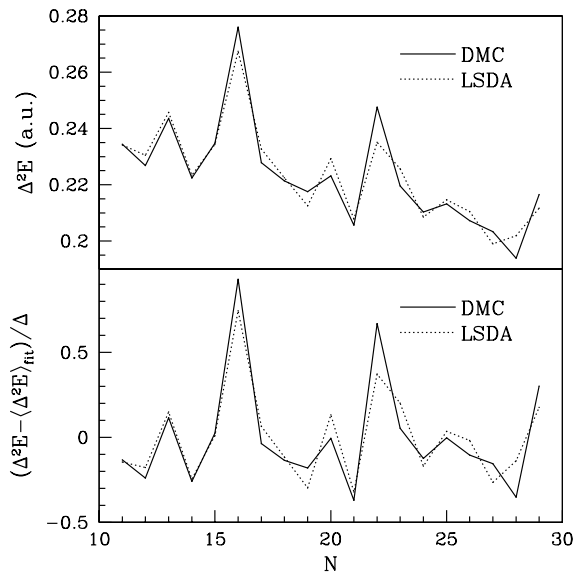}
\caption{
Top: The addition energy from both DMC and LSDA for {\it one} realization of
$V_{\rm ext}$ as a function of the number of electrons on the dot. Note the
large mesoscopic fluctuations in $\Delta^2E$ around the expected overall
decrease. Bottom: Fluctuation in the addition energy after removing the smooth
part, normalized to the mean level spacing $\Delta$. We see that the
fluctuations are of order $\Delta$ and that they are somewhat larger for DMC
results than for LSDA. 
}
\label{fig:Fig2}
\end{figure}

\begin{figure}[t]
\includegraphics[width=3in,clip]{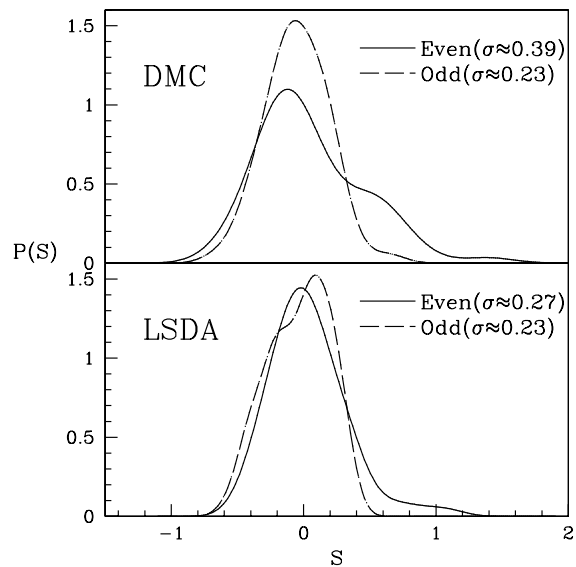}
\caption{
Distribution of the normalized fluctuations in the conductance peak spacing,
$s \!\equiv\!  (\Delta^2E - \langle \Delta^2E \rangle_{\rm fit})/\Delta$ from
DMC (top) and LSDA (bottom) calculations.
The DMC distributions for even (solid) and odd (dashed) $N$ are quite
different while there is less difference for LSDA. The standard
deviation in the different cases, $\sigma$, quantifies this contrast. 
(A sliding Gaussian window of width 0.17 (for even) and 0.1 (for odd) is used
to obtain a smooth curve.)
}
\label{fig:Fig3}
\end{figure}

The ground state spin shows the difference between LSDA and DMC results at
only a coarse level.  To obtain a more detailed understanding, we focus on
the ``spin gap'', $\delta$, which we define as
$\delta = E(S \!=\! 1)-E(S \!=\! 0)$ for even $N$ and
$\delta = E(S \!=\! 3/2)-E(S \!=\! 1/2)$ for odd $N$.
Thus $\delta$ is the amount by which the higher spin state differs in energy
from the lower. Changing the spin of a dot from 0 or 1/2 to a
higher value involves a competition between the single-particle energy cost
and the exchange energy, $-J S(S+1)$, gain.\cite{foot1}  
  
The key result for comparing DMC and LSDA is shown in Fig.~1(b): the
distribution of $(\delta_{\rm DMC}-\delta_{\rm LSDA})/\Delta$. Note, first,
that the distribution is broad (full width $\sim\! 0.4 \Delta$); it is of
order the energy required to flip $S_{\rm GS}$ from $0$ to $1$ for an even
dot, $\Delta-2J$, assuming a realistic value of the exchange parameter
$J \!\sim\! 0.35$ for $r_s \!\sim\! 1.5$. Second, note that the spin gap in
DMC tends to be larger than that in LSDA. This indicates that the strength
of interactions in LSDA is overestimated.
Finally, we have studied this quantity separately for  
the smaller ($N \!=\! 10$-$20$) and larger ($N \!=\! 20$-$30$) dots. We
have not found any size dependence -- results in both ranges of $N$ are
the same as in Fig.~1(b) within our statistical accuracy.
\textit{These observations, together with the results of Fig.~1(a), show that LSDA unduly favors non-trivial spin states}.

The ground state spin distribution has implications for the distribution
of the spacing between Coulomb blockade conductance peaks \cite{denis},
through its relation to $E_{\rm GS}$. In the nearly isolated dot limit, the 
spacing between the Coulomb blockade
conductance peaks is proportional to the the addition energy
\cite{Kouwenhoven97} defined by
   \begin{equation}
\Delta^2E(N)=E_{\rm GS}(N+1)+E_{\rm GS}(N-1)-2E_{\rm GS}(N) \;.
 \label {eq:addn}
   \end{equation}
For noninteracting electrons one would have
\begin{equation}
\Delta^2E(N) = \left\{ \begin{array}{ll}
\epsilon_{N/2} - \epsilon_{N/2-1} &\;\; \mbox{for even } N\\
0 &\;\; \mbox{for odd } N
\end{array} \right. \label{nonint}
\end{equation}
where $\epsilon_i$ are the energies of the single-particle states. Note the
sharply different characteristics of even and odd $N$ dots.
Interactions reduce this strong even/odd effect.

The behavior of $\Delta^2E (N)$ for a particular realization of $V_{\rm ext}$
is presented in the top panel of Fig.~2. Similar qualitative behavior is
observed for all other configurations we studied. An overall decrease
of $\Delta^2E$ with $N$ is expected due to the increase of effective
capacitance of the dot and, hence, the decrease in the classical charging
energy as the dot gets bigger. On top of the mean behavior, we clearly see
strong mesoscopic fluctuations arising from the interplay of electron
interaction and interference effects in the irregular dots. The fluctuations
seem to be slightly larger in the DMC results than in LSDA. To focus on
these fluctuations, we subtract the smooth classical part using a linear
fit, and present the fluctuating part normalized by the mean level spacing
$\Delta$ (which is the natural scale of these mesoscopic fluctuations) in the
lower panel.

To get a more quantitative picture, we plot the distribution of the normalized
addition energy fluctuations in Fig.~3.
\textit{Note the larger difference between the two distributions obtained
from DMC than between those obtained from LSDA.}
In the DMC results, the width of the distribution for even $N$ is
significantly larger
than that for odd dots, and the even $N$ distribution has a long tail
reminiscent of the Wigner surmise for the distribution of 
$\epsilon_{N/2} \!-\! \epsilon_{N/2-1}$ found using random matrix
theory (RMT) [see Eq.~(\ref{nonint})].

Looking at the data for smaller and larger dots separately, we find a strong
trend shown in Fig.~4: \textit{the even/odd effect in the DMC data decreases
significantly as $N$ increases.} If extrapolated to much larger $N$, this trend,
also present in our LSDA
results, contradicts the prediction\cite{denis,gonzalo} of combining RPA
interactions with an RMT treatment of the single particle statistics.

\begin{figure}[t]
\includegraphics[width=3.25in,clip]{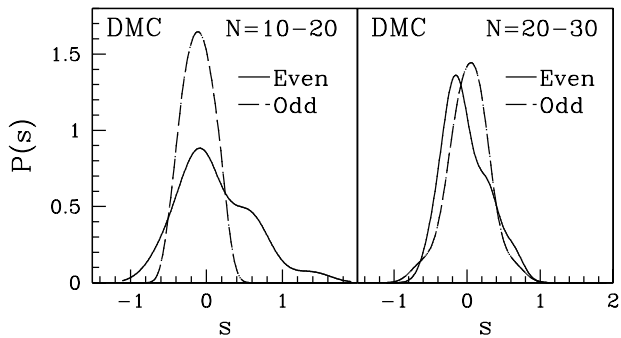}
\caption{
Distribution of the normalized fluctuations in the conductance peak spacing,
from DMC calculations in the range of $N \!=\! 10$-$20$ (left) and
$N \!=\! 20$-$30$ (right).
For smaller $N$, the even/odd effect is stronger:
$\sigma_{\rm even} \!=\! 0.46 \pm 0.08$, while $\sigma_{\rm odd}
\!=\! 0.18 \pm 0.02$.
On the other hand for larger $N$, the even/odd effect is
significantly reduced: $\sigma_{\rm even} \!=\! 0.30 \pm 0.04$ and
$\sigma_{\rm odd} \!=\! 0.25 \pm 0.04$. A similar qualitative trend is found
in the LSDA results, though quantitatively it is somewhat weaker.
}
\label{fig:Fig4}
\end{figure}

In conclusion, we have used quantum Monte Carlo to accurately investigate
the role of interactions in the mesoscopic regime. We find that for
irregular dots with a gas
parameter $r_s \!\sim\! 1.5$ and electron number in the range 10 to 30,
DMC calculations show (1)~mesoscopic fluctuations of the addition energy,
(2)~a substantial probability of non-trivial ground state spin, and (3)~a
significant even/odd effect in the addition energies. In comparison to LSDA,
DMC typically predicts a somewhat larger spin gap;
as a consequence, it has a tendency to find smaller ground state spins
and a somewhat stronger even/odd effect in the addition spectra.
These findings suggest that LSDA, as compared to DMC, in some sense over
predicts the effect of interactions. It is interesting to note that a similar
conclusion concerning overly strong interactions in LSDA was reached in the
large dot regime\cite{denis3} using the Strutinsky analysis scheme. 
 

Acknowledgments: We thank Weitao Yang for helpful
discussions. This work was supported in part by the NSF (grants DMR-0103003
and DMR-0205328).



\begin{thebibliography}{99}
\bibitem{foot3}
In our LSDA calculations, we used the Tanatar-Ceperley parameterized form for
the exchange-correlation energy functional; see Ref.\ \ \onlinecite{Jiang03}
for details.
\bibitem{foot1}
The contribution of the Cooper channel, which should
be included for time reversal invariant systems in a proper
``universal Hamiltonian'' description, is not taken into account here.
\end{thebibliography}
\end{document}